\documentclass[letterpaper]{jpconf}

\usepackage{xspace}
\usepackage{graphicx}
\usepackage{amsmath}
\usepackage{dcolumn}
\newcolumntype{d}{D{.}{.}{-1}}  % not needed in RevTeX
\usepackage{bm}
\usepackage{wasysym}
\usepackage{txfonts}

\newcommand{\nuc}[2]{${}^{#1}$#2}

\renewcommand{\today}{\number\day\space\ifcase\month\or January%
  \or February\or March\or April\or May\or June\or July\or August%
  \or September\or October\or November\or December\fi\space\number\year}
\newcommand{\mc}[3]{\multicolumn{#1}{#2}{#3}}
\newcommand{\mco}[1]{\multicolumn{1}{c}{#1}}

\newcommand{\Be}{$^7$Be}
\newcommand{\Bo}{$^8$B}
\newcommand{\jour}[4]{{#4} {\it #1} {\bf #2} {#3}}

\begin{document}

\title{Radiochemical solar neutrino experiments}

\author{V N Gavrin$^1$ and B T Cleveland$^{2,}$\footnote[3]{Present
address: SNOLAB, PO Box 159, Lively, Ontario P3Y 1M3, Canada}}

\address{$^1$ Institute for Nuclear Research of the Russian Academy of
Sciences, Moscow 117312, Russia}
\address{$^2$ Department of Physics, University of Washington, Seattle
Washington 98195, USA}

\ead{$^1$ gavrin@dionis.iasnet.ru, $^2$ bclevela@snolab.ca}

\begin{abstract}

Radiochemical experiments have been crucial to solar neutrino
research.  Even today, they provide the only direct measurement of the
rate of the proton-proton fusion reaction, $p + p \rightarrow d + e^+
+ \nu_e$, which generates most of the Sun's energy.  We first give a
little history of radiochemical solar neutrino experiments with
emphasis on the gallium experiment SAGE -- the only currently
operating detector of this type.  The combined result of all data from
the Ga experiments is a capture rate of $67.6 \pm 3.7$~SNU.  For
comparison to theory, we use the calculated flux at the Sun from a
standard solar model, take into account neutrino propagation from the
Sun to the Earth and the results of neutrino source experiments with
Ga, and obtain $67.3 ^{+3.9}_{-3.5}$~SNU.  Using the data from all
solar neutrino experiments we calculate an electron neutrino $pp$ flux
of $\phi^{\earth}_{pp}=(3.41^{+0.76}_{-0.77}) \times
10^{10}$/(cm$^2$-s), which agrees well with the prediction from a
detailed solar model of $\phi^{\earth}_{pp}=(3.30 ^{+0.13} _{-0.14})
\times 10^{10}$/(cm$^2$-s).  Four tests of the Ga experiments have
been carried out with very intense reactor-produced neutrino sources
and the ratio of observed to calculated rates is $0.88 \pm 0.05$.  One
explanation for this unexpectedly low result is that the cross section
for neutrino capture by the two lowest-lying excited states in
\nuc{71}{Ge} has been overestimated.  We end with consideration of
possible time variation in the Ga experiments and an enumeration of
other possible radiochemical experiments that might have been.

\end{abstract}

\section{Introduction and a little history}

Our knowledge of neutrinos from the Sun is based on seven experiments:
Homestake, Kamiokande, SAGE, Gallex, GNO, Super-Kamiokande and SNO.
More than half of these are radiochemical experiments.

The detection of neutrinos by use of the inverse $\beta$ decay
reaction was proposed 60 years ago by Bruno Pontecorvo
\cite{Pontecorvo}.  This method of detection, which is the basis for
radiochemical experiments, has played a fundamental role in solar
neutrino investigation.  The idea to use neutrino capture in
\nuc{37}{Cl} to observe the ``undetectable'' new particle proposed by
Wolfgang Pauli was brilliantly realized to observe solar neutrinos by
R. Davis and collaborators in the world-famous experiment at the
Homestake Gold Mine \cite{davis64,davis68,cl98}.  The \nuc{37}{Cl}
experiment was built 4200~m.w.e. (meters of water equivalent)
underground and began to collect data in 1967.  Between 1970--1994,
108 extractions of Ar were made from a tank that contained 615~tons of
C$_2$Cl$_4$.  The number of \nuc{37}{Ar} atoms collected in each run
was measured in a miniature proportional counter.  The result for the
first measured capture rate of solar neutrinos at the Earth was $2.56
\pm 0.23$~SNU.  The SNU unit (defined as 1 neutrino capture per day in
a target that contains $10^{36}$ atoms of the neutrino-absorbing
isotope) was specially introduced by John Bahcall, who had a
fundamental role in the funding of the Cl experiment and the
interpretation of its results, and whose contributions cannot be
overestimated.  Bahcall was the first to fully develop a solar model
that included all the physical parameters needed to calculate the
solar neutrino flux at the Earth.  He worked tirelessly to refine his
calculations and it was the robustness of his solar model that
eventually led all people to understand the significance of the
discrepancy between the result of the Cl experiment and standard solar
model (SSM) predictions.

The discrepancy identified in the Cl experiment attracted the
attention of a significant number of scientists and it soon became
known as ``the solar neutrino problem''.  This problem continued to
bother the mind of scientists for more than 30~years.  Especially
important was the confirmation of the discrepancy by the Kamiokande
experiment \cite{kam96}, a real-time detector of solar neutrinos that
used a completely different method of detection -- electron
scattering, and which began to collect data in 1987.  As a result
there were no doubts that the flux of neutrinos in the high-energy
part of the solar neutrino spectrum was significantly less than the
calculations of the SSM.  Kamiokande, with an analysis threshold of
7~MeV, was sensitive only to the high-energy \nuc{8}{B} neutrinos and
the Cl experiment, whose major response was from the superallowed
analog state at an excitation energy of 5.0~MeV in \nuc{37}{Ar}, was
also mostly sensitive to the \nuc{8}{B} neutrinos.  Another
significant development during this time was the confirmation of the
results of the Bahcall SSM by a solar model independently developed by
Sylvaine Turck-Chi\`{e}ze and collaborators \cite{Turck}.

Despite many attempts, the combination of these two experiments could
not be explained on the basis of solar physics; rather, many
scientists began to believe that it was necessary to reject some of
our old ideas about neutrino properties and to develop new ones.
Conclusive evidence for this suggestion could be obtained by measuring
the low-energy part of the solar neutrino spectrum, which is produced
in reactions that provide the vast majority of the Sun's energy, and
whose flux can be well predicted from the measured solar luminosity
combined with a simple solar model.  The need for experiments
sensitive to low-energy solar neutrinos was recognized shortly after
the first results from the Cl experiment were announced and many
people began to consider radiochemical experiments with low-energy
sensitivity, such as those shown in Table~\ref{Irvine_table}.

\begin{table}[h]
\caption{Radiochemical solar neutrino detectors considered in 1972
{\protect \cite{Irvineconf}} The relative response is given to the
various sources of solar neutrinos.  The mass of target element is the
number of tons required to yield 1 neutrino capture/day from the sum
of the $pp$ and $pep$ reactions.  The relative response and mass were
calculated from the 1972 values of solar flux and used cross sections
that neglected excited states.}
\label{Irvine_table}
\centering
\begin{tabular}{@{} l l r r r r r r @{}}
\br
             &                    & \mc{5}{c}{Relative response (\%)} & Mass  \\ \cline{3-7}
Target       & Product            & $pp$ & $pep$ & \Be & \Bo & CNO    & (tons) \rule{0pt}{2.5ex} \\
\mr
\nuc{87}{Rb} & \nuc{87\rm{m}}{Sr} & 74   &  2    & 21  &  1  &  3     &  32 \\
\nuc{55}{Mn} & \nuc{55}{Fe}       & 67   &  3    & 25  &  1  &  3     & 420 \\
\nuc{71}{Ga} & \nuc{71}{Ge}       & 69   &  2    & 26  &  0  &  3     &  19 \\
\nuc{7}{Li}  & \nuc{7}{Be}        &  0   & 18    & 15  & 51  & 16     &  17 \\
\br
\end{tabular}
\end{table}

From these possibilities attention focused on \nuc{7}{Li}, proposed in
1969 by John Bahcall \cite{BahcallLi}, and on \nuc{71}{Ga}, proposed
in 1965 by Vadim Kuzmin \cite{KuzminGa}.  Because of its high capture
rate, low energy threshold of 233~keV and favorable half-life of
11.4~days, a Ga experiment appeared to be a most attractive
possibility.  The main problems with the Ga experiment were the
acquisition of several tens of tons of the expensive element gallium
and the development of a nearly lossless procedure for the extraction
and purification of \nuc{71}{Ge}.

\section{The Ga experiment}

Laboratory research to develop a gallium experiment began
approximately in 1975.  In the United States this work took place at
Brookhaven National Laboratory under the direction of Ray Davis with
participation of B.~Cleveland, J.~Evans, G.~Friedlander, K.~Rowley,
R.~Stoener from Brookhaven, and W.~Frati and K.~Lande from the
University of Pennsylvania \cite{BNLGa}.  Methods were developed to
extract germanium from liquid gallium metal and from a GaCl$_3$
solution.  After a few years, this group achieved success in
development of these methods and chose the method based on GaCl$_3$
solution.  To carry out the experiment a collaboration was initiated
with a group from the Max Planck Institute at Heidelberg.  Despite
repeated requests and favorable reviews, the Ga experiment was,
however, not funded in the United States.  Rather, a special
subcommittee of the Nuclear Science Advisory Committee recommended
that interested scientists associate themselves with groups in Western
Europe and/or the Soviet Union.  The western European group, called
Gallex, had been formed by the Heidelberg group under the direction of
Till Kirsten when it became apparent that the experiment would not be
funded in the US.

In the Soviet Union, at the Institute for Nuclear Research, laboratory
investigations to develop a gallium experiment began about the same
time in 1975.  It was initially based on a GaCl$_3$ solution, but when
it was learned that Soviet industry could not provide the necessary
radioactive purity in 50~tons of solution, the project was changed to
gallium metal.  Using Davis's idea, the extraction of minute
quantities of \nuc{71}{Ge} from many tons of metallic gallium was
independently developed.  One advantage of metallic Ga is that it is
significantly less sensitive to radioactive impurities.  In 1980 an
installation was built that contained 300~kg of Ga metal.  In addition
to testing the technology, this work also yielded a new limit on the
law of conservation of electric charge \cite{Barabanov}.  By 1985 a
pilot installation containing 7.5~tons of metallic gallium had been
constructed at Troitsk.

The Soviet group built their experiment at the Baksan Neutrino
Observatory in the Caucasus mountains.  The first Ga exposure began in
December 1989 and data collection has continued since that time.
Gallex built their experiment at the Gran Sasso tunnel in Italy and
collected data from 1991-1997.  In 1998 they were reconstituted as the
Gallium Neutrino Observatory (GNO) and they continued operation until
2003 \cite{gno_final}.

\section{SAGE}

In 1986 the Soviet-American collaboration SAGE was officially
established to carry out the gallium solar neutrino experiment at the
Baksan Neutrino Observatory.  The experiment is situated in a
specially built deep underground laboratory where the measured muon
flux is $(3.03 \pm 0.10) \times 10^{-9}/$(cm$^2$ s).  It is located
3.5~km from the entrance of a horizontal adit excavated into the side
of a mountain.  The rock gives an overhead shielding equivalent to
4700~m of water and reduces the muon flux by a factor of $10^7$.

The mass of gallium used in SAGE at the present time is about
50~tonnes.  It is in the form of liquid metal and is contained in 7
chemical reactors.  A measurement of the solar neutrino capture rate
begins by adding to the gallium a stable Ge carrier.  The carrier is a
Ga-Ge alloy with a known Ge content of approximately $350~\mu$g and is
distributed equally among all reactors.  The reactor contents are
stirred thoroughly to disperse the Ge throughout the Ga mass.  After a
typical exposure interval of four weeks, the Ge carrier and
\nuc{71}{Ge} atoms produced by solar neutrinos and background sources
are chemically extracted from the Ga using procedures described in
\cite{prc,jetp}.  The final step of the chemical procedure is the
synthesis of germane (GeH$_4$), which is used as the proportional
counter fill gas with an admixture of (90--95)\% Xe.  The total
efficiency of extraction is the ratio of mass of Ge in the germane to
the mass of initial Ge carrier and is typically $(95 \pm 3)\%$.  The
systematic uncertainty in this efficiency is 3.4\%, mainly arising
from uncertainties in the mass of added and extracted carrier.  The
proportional counter is placed in the well of a NaI detector that is
within a large passive shield and is counted for a typical period of
4--6 months.

Based on criteria described in \cite{prc}, a group of events is
selected from each extraction that are candidate \nuc{71}{Ge} decays.
These events are fit to a maximum likelihood function \cite{max_like},
assuming that they originate from an unknown but constant-rate
background and the exponentially decaying rate of \nuc{71}{Ge}.  A
single run result has little significance because of its large
statistical uncertainty.

The global best fit capture rate for all SAGE data from January 1990
through December 2005 (139 runs and 264 separate counting sets) is
$66.2 ^{+3.5}_{-3.4}$~SNU, where the uncertainty is statistical only.
If one considers the $L$-peak and $K$-peak data separately, the
results are $67.6 ^{+5.5}_{-5.3}$~SNU and $65.5 ^{+4.7}_{-4.5}$~SNU,
respectively.  The agreement between the two peaks serves as a strong
check on the robustness of the event selection criteria.  The
systematic effects fall into three main categories: those associated
with extraction efficiency, with counting efficiency and with
backgrounds.  For a complete description of these effects see
\cite{prc}.  Including all uncertainties, our overall result is
$66.2 ^{+3.5}_{-3.4} \text{ (stat)} ^{+3.8}_{-3.4} \text{
(syst)}$~SNU.  If we combine the SAGE statistical and systematic
uncertainties in quadrature, the result is $66.2 ^{+5.2}_{-4.8}$~SNU.

The final result from 123 runs in the Gallex and GNO experiments is
$69.3 \pm 5.5 \text{ (stat + syst)}$~SNU \cite{gno_final}.  The
weighted combination of all the Ga experiments, SAGE, Gallex and GNO,
is thus
\begin{equation}
67.6 \pm 3.7 \text{ SNU.  \quad \quad Present Ga experiment result.}
\end{equation}

It was very good that for many years there were two Ga experiments
operating at the same time and it is indeed unfortunate that the GNO
experiment was terminated for non-scientific reasons.

\section{Source experiments}

\begin{figure}
\centering
\includegraphics*[width=0.6\hsize,bb= 20 31 442 280]{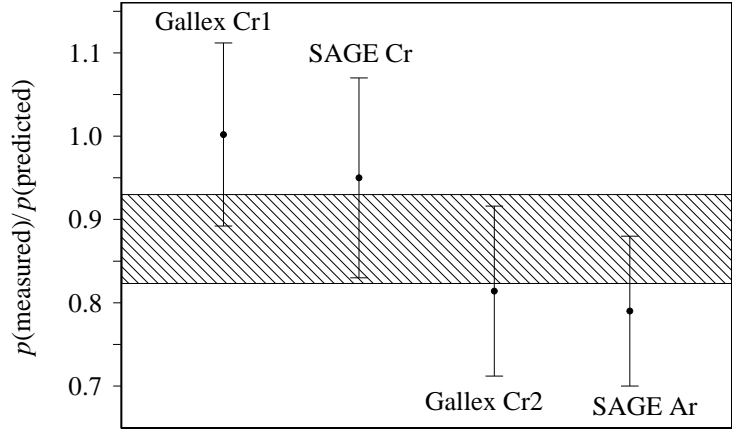}\hspace{0.1\hsize}%
\begin{minipage}[b]{0.2\hsize}\caption{Results of all neutrino source
experiments with Ga.  The hashed region is the weighted average of the
four experiments.  See {\protect \cite{ArPRC} for
details.}\\\\}\end{minipage}
\label{src_res}
\end{figure}

The experimental procedures of both Ga experiments, including the
chemical extraction, counting and analysis techniques, have been
checked by exposing the gallium target to reactor-produced neutrino
sources whose activity was close to 1~MCi.  Gallex has twice used
\nuc{51}{Cr} sources to irradiate their entire target; SAGE has
irradiated about 25\% of their target with a \nuc{51}{Cr} source and
an \nuc{37}{Ar} source \cite{Haxton,Gavrin_preprint}.  The results,
expressed as the ratio $R$ of the measured \nuc{71}{Ge} production
rate to that expected due to the source strength, are shown in
Figure~\ref{src_res}.  The weighted average value of the ratio for the
four experiments is $R = 0.88 \pm 0.05$, more than two standard
deviations less than unity.

Since other auxiliary tests, especially the \nuc{71}{As} experiment of
Gallex, have given great confidence in the knowledge of the various
efficiencies in the Ga experiments, the combined result of these
source tests should not be considered to be a measurement of the
entire throughput of the Ga experiments.  Rather, we believe that,
although not statistically conclusive, the combination of these
experiments suggests that the predicted rates may be overestimated.
The most likely hypothesis \footnote{For an alternative explanation,
based on transitions to sterile neutrinos, see \cite{Giunti}.} is that
the cross sections for neutrino capture to the lowest two excited
states in \nuc{71}{Ge}, both of which can be reached using either
\nuc{51}{Cr} or \nuc{37}{Ar} sources, have been overestimated
\cite{Haxtoncross}.  If the contribution of these two excited states
to the predicted rate is set to zero, then $R = 0.93 \pm 0.05$,
reasonably consistent with unity.  A new experiment with a
considerably higher rate from the neutrino source is needed to settle
this question.

As a side note, during the time of the SAGE \nuc{37}{Ar} source
experiment, which used 26~tonnes of Ga, solar neutrino extractions
were also made from the remaining 22~tonnes of Ga.  Since the SAGE
counting system was filled with samples from the \nuc{37}{Ar} source,
we transported the \nuc{71}{Ge} extracted from the solar runs to Gran
Sasso, where GeH$_4$ was synthesized and the samples were counted in
the GNO counting system.  The combined result of six such solar runs
was $64 ^{+24}_{-22}$~SNU \cite{BNOLNGS}, in excellent agreement with
the overall result of the Ga experiments.

\section{Comparison of gallium result to predictions of standard solar model}

The capture rate $R_i$ of component $i$ of the solar neutrino spectrum
is given in a radiochemical experiment by
\begin{equation}
R_i = \phi^{\astrosun}_i \langle P^{ee}_i\rangle \langle\sigma_i\rangle
\end{equation}
where $\phi^{\astrosun}_i$ is the amplitude of the flux from this
solar component at the production point in the Sun, $\langle
P^{ee}_i\rangle$ is the integral over the solar spectrum of the
probability of survival of the electron neutrino during its travel
from where it is produced in the Sun to where it is detected at the
Earth, and $\langle\sigma_i\rangle$ is the integral of the cross
section for neutrino capture over the spectrum at the Earth.  The
physical origin for the reduction of the electron component of the
solar neutrino flux is the now well-established mechanism of MSW
neutrino oscillations \cite{MSW}.

\begin{table}

\caption{Factors needed to compute the capture rate in \nuc{71}{Ga}
solar neutrino experiments.  The units of flux are $10^{10}(pp)$,
$10^{9}({\rm ^7Be})$, $10^{8}(pep, {\rm ^{13}N, ^{15}O})$, $10^{6}
({\rm ^8B, ^{17}F})$, and $10^{3}(hep)$ ${\rm cm^{-2}s^{-1}}$.  The
uncertainty values are at 68\% confidence.}

\def\arraystretch{1.4}
\label{rate_table}
\centering
\begin{tabular*}{1.00\hsize}{@{} l @{\extracolsep{\fill}} c c c c c c @{}}
\br
Spectrum    & \mc{2}{c}{Flux $\phi^{\astrosun}_i$}&                                 & \mco{$\langle\sigma_i\rangle$}     & \mc{2}{c}{Capture rate $R_i$ (SNU)}   \\ \cline{2-3} \cline{6-7}
comp. $i$   &    BP04                      &BP04+ & \mco{$\langle P^{ee}_i\rangle$} & \mco{($10^{-46}\text{ cm}^2)$}     & BP04                           &BP04+ \\
\mr
$pp$        & $ 5.94 (1^{+0.01}_{-0.01}) $ & 5.99 & $ 0.555 (1^{+0.038}_{-0.040}) $ & $    11.75 (1^{+0.024}_{-0.023}) $ & $ 38.7 (1^{+0.046}_{-0.047}) $ & 39.1 \\
$pep$       & $ 1.40 (1^{+0.02}_{-0.02}) $ & 1.42 & $ 0.517 (1^{+0.033}_{-0.034}) $ & $   194.4  (1^{+0.17 }_{-0.024}) $ & $ 1.41 (1^{+0.17 }_{-0.046}) $ & 1.43 \\
\nuc{7}{Be} & $ 4.86 (1^{+0.12}_{-0.12}) $ & 4.65 & $ 0.537 (1^{+0.036}_{-0.037}) $ & $    68.22 (1^{+0.070}_{-0.023}) $ & $ 17.8 (1^{+0.14 }_{-0.13 }) $ & 17.0 \\
\nuc{13}{N} & $ 5.71 (1^{+0.37}_{-0.35}) $ & 4.06 & $ 0.539 (1^{+0.036}_{-0.038}) $ & $    56.86 (1^{+0.099}_{-0.023}) $ & $ 1.75 (1^{+0.38 }_{-0.35 }) $ & 1.24 \\
\nuc{15}{O} & $ 5.03 (1^{+0.43}_{-0.39}) $ & 3.54 & $ 0.531 (1^{+0.035}_{-0.036}) $ & $   107.2  (1^{+0.13 }_{-0.023}) $ & $ 2.86 (1^{+0.45 }_{-0.39 }) $ & 2.02 \\
\nuc{17}{F} & $ 5.91 (1^{+0.44}_{-0.44}) $ & 3.97 & $ 0.531 (1^{+0.035}_{-0.036}) $ & $   107.8  (1^{+0.13 }_{-0.023}) $ & $ 0.03 (1^{+0.46 }_{-0.44 }) $ & 0.02 \\
\nuc{8}{B}  & $ 5.79 (1^{+0.23}_{-0.23}) $ & 5.26 & $ 0.374 (1^{+0.044}_{-0.039}) $ & $ 21580    (1^{+0.32 }_{-0.15 }) $ & $ 4.67 (1^{+0.40 }_{-0.28 }) $ & 4.25 \\
$hep$       & $ 7.88 (1^{+0.16}_{-0.16}) $ & 8.04 & $ 0.347 (1^{+0.061}_{-0.054}) $ & $ 66300    (1^{+0.33 }_{-0.16 }) $ & $ 0.02 (1^{+0.37 }_{-0.23 }) $ & 0.02 \\
\mr
Total       &                              &      &                                 &                                    & $ 67.3 ^{+3.9}_{-3.5}        $ & 65.1 \\
\br
\end{tabular*}
\end{table}

Values of $\phi^{\astrosun}_i, \langle P^{ee}_i\rangle$ and $\langle
\sigma_i\rangle$ are given for each neutrino component in
Table~\ref{rate_table}.  The fluxes are from two solar models with
differing composition \cite{BAH04}.  The other quantities were
calculated assuming three-neutrino mixing to active neutrinos with
parameters $\Delta \text{m}^2_{12}=(7.92 \pm 0.36) \times 10^{-5}
\text{ eV}^2, \theta_{12}=34.1^{+1.7}_{-1.5} \text{ degrees and }
\theta_{13}=5.44^{+2.79}_{-5.44} \text{ degrees} \, \cite{Bari}$.  The
approximate formulae given in \cite{Barger} were used for the survival
probability $P^{ee}_i(E)$.  Since radiochemical experiments average
over a long exposure interval, regeneration in the Earth was
neglected.  The cross sections $\sigma(E)$ were taken from
\cite{gacross} but were modified to delete the effect of the lowest
two excited states in \nuc{71}{Ge} according to the results of the
neutrino source experiments as given in the previous section.  The
neutrino spectra were taken from \cite{gacross} ($pp$ and CNO),
\cite{clcross} (\nuc{8}{B}) and \cite{website} ($hep$).

There is excellent agreement between the calculated ($67.3
^{+3.9}_{-3.5}$~SNU) and observed ($67.6 \pm 3.7$~SNU) capture rates
in \nuc{71}{Ga}.

\section{The $\bm{\lowercase{pp}}$ neutrino flux}

One of the main purposes of the Ga experiment is to provide
information that leads directly to the experimental determination of
the flux of $pp$ neutrinos at the Earth.  In this Section we will
assume the Sun is generating energy by the $pp$ cycle, and not
dominantly by the CNO cycle, and will derive the present best value
for the $pp$ flux directly from the results of neutrino experiments.

To obtain the $pp$ flux we begin with the combined capture rate from
the SAGE and GALLEX/GNO experiments given above of $67.6 \pm 3.7$~SNU.
This rate is the sum of the rates from all the components of the solar
neutrino flux, which we denote by [$pp$+$^7$Be+CNO+$pep+^8$B$|$Ga].
(We ignore the $hep$ contribution.)

The only one of these flux components that is known from direct
experiment is the \nuc{8}{B} flux, measured by SNO to be [$^8$B$|$SNO]
= $(1.68 \pm 0.11) \times 10^6$ electron neutrinos/(cm$^2$-s)
\cite{SNO05} at the Earth.  We multiply this flux by the cross section
for \nuc{8}{B} given in Table~\ref{rate_table} and find that the
contribution to the Ga experiment is [$^8$B$|$Ga] =
$3.7^{+1.2}_{-0.7}$ SNU.  Subtracting this measured value from the
total Ga rate gives [$pp+^7$Be+CNO+$pep|$Ga] = $64.0^{+3.7}_{-3.9}$
SNU.

The measured capture rate in the Cl experiment is
[$^7$Be+CNO+$pep+^8$B$|$Cl] = $2.56 \pm 0.23$ SNU \cite{cl98}.  In a
manner analogous to Ga we can calculate the cross section for
\nuc{8}{B} neutrinos on \nuc{37}{Cl}, including the suppression
factor, to be $1.02 (1 \pm 0.046) \times 10^{-42}$ cm$^2$.  We
multiply this by the flux measured by SNO and deduce that the
contribution of \nuc{8}{B} to the Cl experiment is [$^8$B$|$Cl] =
$1.72 \pm 0.14$ SNU.  Subtracting this component from the total leaves
[$^7$Be+CNO+$pep|$Cl] = $0.84 \pm 0.27$ SNU, all of which is due to
neutrinos of medium energy.

We assume the Sun is generating its energy via the $pp$ cycle so these
medium-energy neutrinos are dominated by \nuc{7}{Be}.  We can thus
make the approximation that [$^7$Be+CNO+$pep|$Ga] =
[$^7$Be+CNO+$pep|$Cl] $\times$ cross section for \nuc{7}{Be} on
Ga/cross section for \nuc{7}{Be} on Cl = ($0.84 \pm 0.27) \times [71.9
(1 + ^{+0.07} _{-0.03})]/[2.40 (1 \pm 0.02)] = 23.9^{+7.9}_{-7.6}$
SNU.  There is an additional error due to the approximation used,
which is estimated to be 10\%, giving the result [$^7$Be+CNO+$pep|$Ga]
= $23.9^{+8.1}_{-8.0}$ SNU.

We subtract this contribution from the rate given above and get the
result for the measured $pp$ rate in the Ga experiment [$pp|$Ga] =
[$pp+^7$Be+CNO+$pep|$Ga] - [$^7$Be+CNO+$pep|$Ga] =
$40.1^{+6.6}_{-9.0}$ SNU.  Dividing this capture rate by the cross
section for capture of $pp$ neutrinos of $11.8 (1^{+0.024}_{-0.023})
\times 10^{-46}$ cm$^2$ gives the measured electron neutrino $pp$ flux
at Earth of $\phi^{\earth}_{pp}=(3.41^{+0.76}_{-0.77}) \times
10^{10}$/(cm$^2$-s).  The major component of the error in this $pp$
flux measurement is due to the poor knowledge of the medium-energy
neutrinos which was inferred from the Cl experiment.

For comparison, the standard solar model calculates the $pp$ flux
produced in the Sun to be $\phi^{\astrosun}_{pp}=5.94 (1 \pm 0.01)
\times 10^{10}$/(cm$^2$-s) \cite{BAH04}
\footnote{The error here is only 1\% because the measured solar
luminosity was used in this calculation.}
.  If we multiply this rate by the average survival probability for
$pp$ neutrinos, which from Table~\ref{rate_table} is $0.555
(1^{+0.038}_{-0.040})$, we obtain a $pp$ flux at the Earth of
$\phi^{\earth}_{pp}=(3.30 ^{+0.13}_{-0.14})\times 10^{10}$/(cm$^2$-s),
in excellent agreement with the value determined above from solar
neutrino experiments.

In the future it will be possible to reduce the error in this flux
measurement when there are new experiments that directly measure the
\nuc{7}{Be} flux, as anticipated by Borexino and KamLAND, and the CNO
flux, as anticipated by SNO+.  The dominant error should eventually be
due to the inaccuracy of the Ga measurement itself.

\section{Is the neutrino capture rate in Ga constant?}

Short-term variations in the Gallex-GNO rate with periods from 15 days
to a few 100~days have been considered by Pandola
\cite{Pandola_time_var} and Sturrock {\it et al.} \cite{Sturrock_ga}.
Pandola's analysis finds no variability but Sturrock {\it et al.} see
evidence for variation if one considers the Gallex and GNO data sets
separately.

\begin{figure}
\centering
\includegraphics*[width=0.526\hsize,bb= 36 39 439 286]{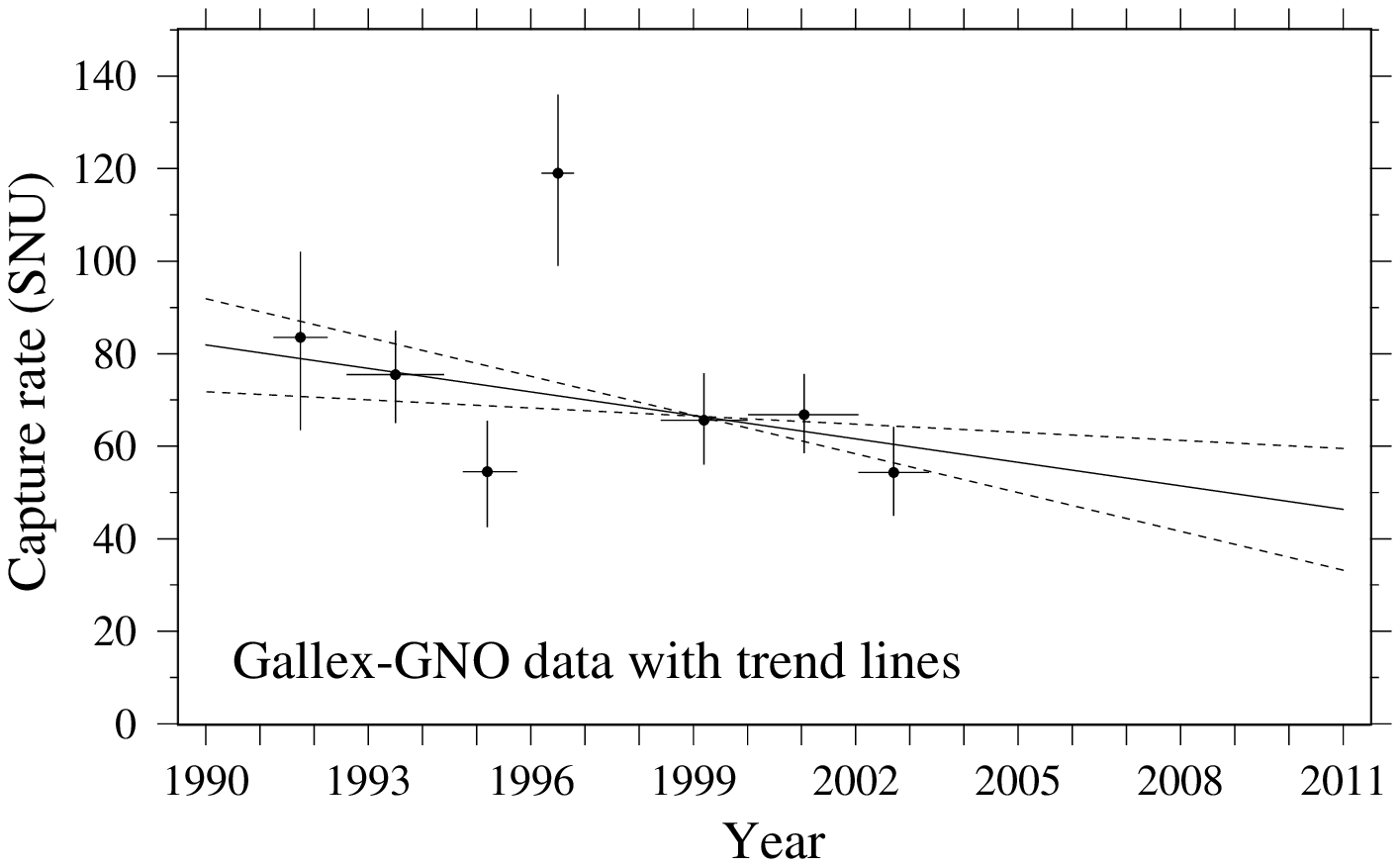}%
\includegraphics*[width=0.47\hsize,bb= 79 39 439 286]{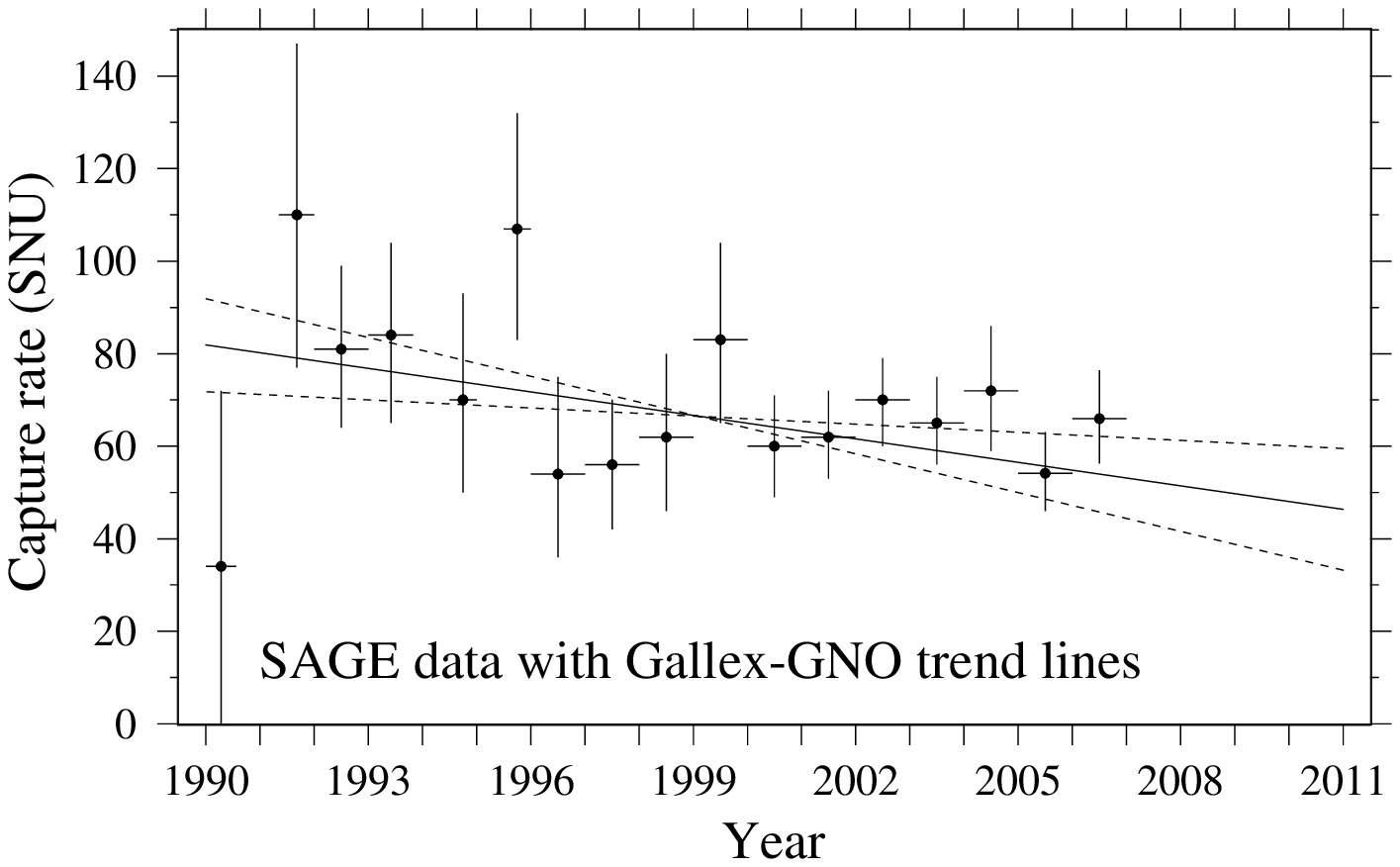}
\caption{Gallex-GNO results (left panel) and SAGE results (right
panel) vs time.  See text for further explanation.}
\label{ga_periods}
\end{figure}

The possibility of variability over longer time periods has been
considered by several authors \cite{Strumia,nu2004}.  In a plot of
the data there appears to be a difference between early and late time
periods, which gives a visual hint of a long-term decrease, as
illustrated in Figure~\ref{ga_periods}.  The Gallex-GNO data is shown
on the left of this Figure where the data have been grouped by the
experimenters into 7 intervals.  The SAGE data, divided into intervals
of one calendar year, is shown on the right of
Figure~\ref{ga_periods}.  The average rate prior to 1997 is higher in
both experiments than in the data after 1997.

If one assumes the rate in Gallex-GNO varies linearly in time then the
best fit gives \cite{gno_final}
\begin{equation}
\text{Capture rate} = 82 \pm 10 - (1.7 \pm 1.1) \times [\text{t(year) - 1990}].
\end{equation}
These trend lines are plotted for both experiments in
Figure~\ref{ga_periods} and there is reasonably good visual agreement
with the measured data.

When examined quantitatively, however, the evidence for long-term
variability becomes less convincing.  A $\chi^2$ test applied to the
Gallex-GNO data with (without) the assumed time variation yields
$\chi^2$/dof = 10.8/5 (13.2/6), prob.\ = 5.6\% (4.0\%), i.e., the fit
to both the time-varying rate and to a constant rate is more or less
equally bad.  For the SAGE data the fit to a constant rate gives
$\chi^2$/dof = 11.7/16, prob.\ = 76\%, whereas the fit to the central
Gallex-GNO trend line yields $\chi^2$/dof = 11.4/17, prob.\ = 83\%,
i.e., the fit to both rate hypotheses is quite good.  At the present
time we cannot differentiate between these two hypotheses, but it
should become possible to do so with considerable additional data.

Up to now it is not known if this apparent variability is a
statistical fluctuation or an indication of a real effect, such as has
been considered by Pulido {\it et al.} \cite{Pulido}.

\section{Other radiochemical experiments}

Several other radiochemical experiments to measure solar neutrinos
have been developed to various degrees.  These include \nuc{127}{I}
$\rightarrow$ \nuc{127}{Xe} \cite{HaxtonI,iodine} and \nuc{81}{Br}
$\rightarrow$ \nuc{81}{Kr} \cite{bromine} experiments that would in
many ways be similar to the \nuc{37}{Cl} experiment, and a
\nuc{97}{Mo} $\rightarrow$ \nuc{97}{Tc} \cite{mo-tc} experiment that
could measure the long-term history of the \nuc{8}{B} solar neutrino
flux.  Although very considerable efforts were expended in the United
States on the \nuc{127}{I} and \nuc{97}{Mo} experiments, they were
never brought to fruition, mainly because of a lack of funding.

The \nuc{7}{Li} $\rightarrow$ \nuc{7}{Be} experiment has continued to
be pursued in Russia.  Methods for the efficient extraction of Be from
metallic Li have been proven \cite{GavrinLi} and an experiment could,
in principle, be built \cite{Kopylov}.

At the present time interest in radiochemical experiments has greatly
decreased and it is only direct-counting experiments that are under
development.  Nonetheless, the radiochemical experiments stimulated
great interest in the solar neutrino problem, which led to the
real-time experiments Super-Kamiokande, SNO and KamLAND.

\ack

We thank Thomas Bowles for leading the American side of the SAGE
collaboration for many years and for organizing this excellent
conference.  We are grateful to our SAGE colleagues for their hard
work and perseverance.  We wish to thank Victor Matveev, Valery
Rubakov and Albert Tavkhelidze of the Institute for Nuclear Research
RAS, Russia for their vigorous and continuous support.

\end{document}